\documentstyle[aps,prl,twocolumn,epsfig]{revtex}

\def\ps#1{\begin{center}\leavevmode\hbox{\epsfxsize=2.7in\epsfysize=3.3in\epsfbox{#1}}\end{center}}

\def\psss#1{\begin{center}\leavevmode\hbox{\epsfxsize=2.7in\epsfysize=3.1in\epsfbox{#1}}\end{center}}

\def\ell{i}

\begin{document}
\title{The Anderson prescription for surfaces and impurities}
\author{K. Tanaka and F. Marsiglio}
\address{Department of Physics, University of Alberta,
Edmonton, Alberta, Canada T6G 2J1}
\date{\today}

\twocolumn[\hsize\textwidth\columnwidth\hsize\csname@twocolumnfalse\endcsname
\maketitle
\begin{abstract}
We test the Anderson prescription \cite{anderson59}, a BCS formalism 
for describing superconductivity in inhomogeneous systems,
and compare results with those obtained from the Bogoliubov-de Gennes 
formalism, using the attractive Hubbard model
with surfaces and nonmagnetic impurities.
The Anderson approach captures the essential features of
the spatial variation of the gap parameter and electron density
around a surface or an impurity over a wide range of parameters. 
It breaks down, however,
in the strong-coupling regime for a weak impurity potential.
\end{abstract}
\pacs{74.20.Fg,74.80.-g,73.20.-r}
]
\narrowtext

\makeatletter
\global\@specialpagefalse
\def\@oddhead{\hfill Alberta Thy 01-00}
\let\@evenhead\@oddhead
\makeatother

In microscopic treatments of
inhomogeneity effects in superconductors, impurities are often
averaged over in some manner \cite{anderson59,abrikosov59}. In the 
last decade, partly because of the increased computational resources now 
available, and partly
because of the technical advances which allow small particle fabrication and
single atom manipulation \cite{physics_today}, the role of inhomogeneous 
effects 
in superconductors has received more wide-spread attention \cite{rainerbook}. 
One of the theoretical frameworks for
addressing these questions is the Bogoliubov-de Gennes (BdG) formalism 
\cite{degennes66}. This formalism allows one to answer questions regarding 
surfaces, interfaces, and
impurity effects at a level of detail not previously addressed. At present,
however, the scope of problems for which one can compute results accurately
is limited by computer resources, since matrices whose dimension grows with
system size require full diagonalization and should be solved 
self-consistently.  On the other hand, 
the Anderson prescription \cite{anderson59}, first presented to 
examine the impact of impurities on superconductivity, 
is a BCS formalism 
for an inhomogeneous system and requires {\it one} diagonalization
of the single-particle problem.
The purpose of this paper
is to examine the limits of applicability of the Anderson approach, as
compared to the BdG formalism. We first summarize the two approaches,
and then follow with some concrete examples, utilizing surfaces and
impurities as sources of inhomogeneities.

We find that the Anderson prescription works very well for surfaces
and in some regimes for a single impurity.  It breaks down for strong
coupling with weak impurity scattering.

For our purposes we adopt a convenient model to describe s-wave 
superconductivity,
the attractive Hubbard Hamiltonian (the results of our study will presumably
apply to d-wave or other symmetry states): 
\begin{eqnarray}
H -\mu N_e &=& -\,\sum_{i,\delta \atop \sigma}t_\delta\,
(a_{i+\delta,\sigma}^\dagger a_{i\sigma} +{\rm h.c.})\nonumber\\  
&&- \sum_{i, \sigma} (\mu - \epsilon_i) n_{i\sigma} - 
|U| \sum_{i} n_{i\uparrow} n_{i\downarrow}.
\label{hamil}
\end{eqnarray}
Here, $a_{i\sigma}^\dagger$ ($a_{i\sigma}$) creates (annihilates) an electron
with spin $\sigma$ at site $i$
and $n_{i\sigma}$ is the number operator for an electron with spin $\sigma$
at site $i$.
The $t_\delta$ is the hopping rate of electrons from one site to a neighbouring
site (often nearest neighbours only are included, and we will adopt this
model here),
and $|U|$ is the attractive coupling strength between electrons on the same 
site.
As usual, this attraction is justified in terms of an electron-phonon coupling,
where retardation effects are unimportant, as is the case with many 
conventional 
superconductors. The second term includes 
the chemical potential $\mu$
and the impurity potential at site $i$, $\epsilon_i$. 
We assume that impurities act
to raise or lower single site energy levels. 
It is worth noting that
impurity effects can certainly enter in other ways.
For example if an impurity
occupies one of the sites, undoubtedly the hopping amplitude 
to and from that site will also be altered, as will the interaction between
two electrons occupying the same orbital on that site. 
In many studies (see, for example, Ref. \cite{randeria98}), the 
$\epsilon_i$ are randomly distributed with some
probability distribution, and then the results are averaged, to reflect
the fact that we generally have no control over the precise distribution
of impurities in the bulk. However, in systems where a single impurity
can be added to the surface, for example (see, eg., Ref. \cite{hudson99}),
we would want to study this model with only one impurity, at a particular
site (in this case, on the surface).

Equation (\ref{hamil})
also allows us the freedom to choose periodic boundary conditions (PBC) (to recover
well-known results) or `open' boundary conditions (OBC). The latter are natural in a
tight-binding context; they require no assumptions about the order parameter,
for example. `Open' here simply means that electrons cannot hop beyond the 
surface.  Here again more sophisticated boundary effects could be 
included --- for
example, in a real system the hopping integral at the surface will no 
doubt differ from that in the bulk, but we leave aside these finer points.

The BdG equations are obtained by defining an effective Hamiltonian,
with effective potentials \cite{degennes66}. By diagonalizing this
effective Hamiltonian through the generalized Bogoliubov-Valatin
transformation \cite{degennes66},
one arrives at the two BdG equations \cite{hirsch92}:
\begin{eqnarray}
E_n u_n(\ell) = \phantom{-} \sum_{\ell^\prime}
A_{\ell \ell^\prime}u_n(\ell^\prime) + V_\ell u_n(\ell) + \Delta_\ell v_n(\ell)
\label{bdgu}\\
E_n v_n(\ell) = -\sum_{\ell^\prime} A_{\ell \ell^\prime}v_n(\ell^\prime)
- V_\ell v_n(\ell) + \Delta^\ast_\ell u_n(\ell)
\label{bdgv}
\end{eqnarray}
where
\begin{equation}
A_{\ell \ell^\prime} = -t \sum_\delta \biggl(
\delta_{\ell^\prime, \ell - \delta} + \delta_{\ell^\prime, \ell + \delta}
\biggr)
-\delta_{\ell \ell^\prime} \biggl(\mu - \epsilon_\ell \biggr).
\label{bdgaux}
\end{equation}
The self-consistent potentials, $V_\ell$, and $\Delta_\ell$, are given
by
\begin{eqnarray}
\Delta_{\ell} = |U| \sum_n u_n(\ell) v_n^\ast(\ell) (1 - 2f_n)
\phantom{aaaaaaaaa}
\label{delpot} \\
V_\ell = -|U| \sum_n \biggl[
|u_n(\ell)|^2 f_n + |v_n(\ell)|^2 (1 - f_n) \biggr].
\label{harpot}
\end{eqnarray}
We use the index $n$ to label the eigenvalues (there are $2N$ of them),
the index $\ell$ to label
the sites (1 through N), and the composite eigenvector is
given by $\biggl({u_n \atop v_n} \biggr)$, of total length $2N$. The sums
in Eqs. (\ref{delpot},\ref{harpot}) are over positive eigenvalues only.
The $f_n$ is the Fermi function, with argument $\beta E_n$, where
$\beta \equiv {1/k_BT}$, with $T$ the temperature. The single
site electron density, $n_\ell$, is given, through Eq. (\ref{harpot}),
by $V_\ell = -|U|n_\ell/2$.

The equations (\ref{delpot},\ref{harpot}) for the effective potentials were 
determined through a variational principle
so that the effective Hamiltonian allows fluctuations
in any number of mean fields \cite{degennes66}. As written, there are two
possible mean field potentials, the {\em Hartree potential}, $V_\ell$, and the
{\em pair potential}, $\Delta_{\ell}$,
from which the ground state energy and other properties may be obtained.

An alternate prescription was originally proposed by 
Anderson \cite{anderson59},
whereby one first solves for the eigenvalues and eigenstates of the 
`non-interacting' problem, i.e.,
\begin{equation}
E^0_n w_n(\ell) = \sum_{\ell^\prime}
A_{\ell \ell^\prime}w_n(\ell^\prime).
\label{andw}
\end{equation}
Using the unitary matrix, $U_{\ell n}$, for a basis which diagonalizes
the single-particle Hamiltonian, one can determine
the transformed electron-electron interaction:
\begin{equation}
V_{nm,n^\prime m^\prime} = -|U| \sum_{\ell} U^\ast_{\ell n} U^\ast_{\ell m}
U_{\ell n^\prime} U_{\ell m^\prime},
\label{interaction}
\end{equation}
which now mediates the (generally off-diagonal) electron-electron interaction.
The gap and number equations are obtained just as in BCS theory, except that
now the label is not the wave vector {\bf k}, but rather some quantum number 
$n$, which simply enumerates the single particle eigenvalues \cite{tanaka00}.
From the solution, one can obtain the ground state energy
and, by transforming back to space coordinates, site-dependent 
quantities.

The simplest origin of gap inhomogeneity in a superconductor is the surface.
The presence of surfaces beyond which electrons are unable to move
yields a gap parameter (i.e., pair potential)
which can exhibit a variety of behaviour
near the surface. Traditionally in Ginzburg-Landau treatments the
gap function is given a priori a boundary condition \cite{degennes66};
here the behaviour near a boundary (or impurity) is a derived quantity,
i.e. as the solution to the BdG (or Anderson) equations. 

The advantage (for the Anderson approach) of examining the impact
of surfaces on the gap parameter is that an analytical solution
exists for a simple tight-binding model \cite{trugman}. The eigenstates
for a chain \cite{chain} of length N, with lattice spacing $a$ and 
nearest-neighbour hopping $t$, are
\begin{equation}
a_{k \sigma} = \sqrt{{2 \over N + 1}} \sum_i\,\sin{(kR_i)}\,a_{i \sigma},
\label{obcstates}
\end{equation}
and the eigenenergies are
\begin{equation}
E_k^{(0)} = -2t\cos{(ka)}\;, \quad ka = {\pi n \over N+1}\;,
\label{obcenergies}
\end{equation}
where $n = 1, 2, ..., N$.
We use these analytical results in the Anderson approach (making it not
significantly more difficult than BCS theory), while in the BdG approach
these analytical solutions are not particularly helpful.

\begin{figure}
\ps{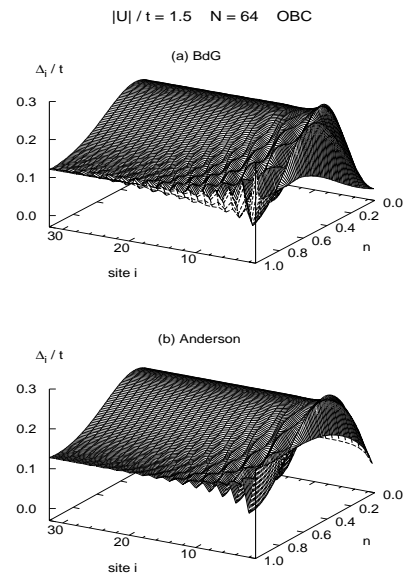}
\caption{Gap parameter $\Delta_i$ as a function of site number $i$
and electron density $n$, for $N=64$ with OBC
and $|U| = 1.5\,t$. The BdG and Anderson results are shown in (a) and (b),
respectively.}
\end{figure}

In Fig.~1 we show the gap parameter $\Delta_i$ 
as a function of site number $i$, for electron
density $n$ ranging from half-filling to zero.
The chain length is $64$ sites, and we have used OBC
and the coupling strength $|U| = 1.5\,t$.
Here $\Delta_i$ is shown for half the chain length (from $i=1$ to $32$): 
the gap parameter is symmetric about the middle.  In Fig.~1(a) we plot
the result from the BdG equations, while in Fig.~1(b) we show the
corresponding results from the Anderson prescription. The first
thing to note is that in either case the behaviour near
the surface is markedly different as a function of electron density.
At half-filling the gap parameter actually peaks at the surface, with
several `Friedel-like' oscillations ensuing towards the center of the
sample, while at low fillings the gap parameter is much smoother by
comparison. A comparison of the two figures shows quantitative differences,
but overall, qualitatively they are very similar. It is evident that
the Anderson prescription captures the essence of the BdG results
remarkably well.

\begin{figure}
\psss{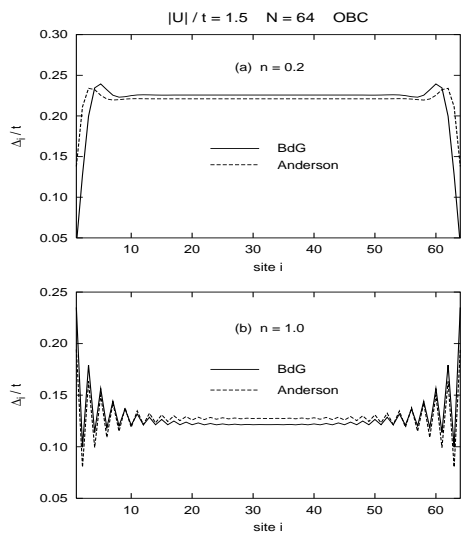}
\caption{Cross sections of $\Delta_i$ shown in Fig.~1 at density
(a) $n=0.2$ and (b) $n=1.0$.  The BdG and Anderson results are now plotted
together, in solid and dashed lines, respectively.  It can be clearly seen
that the Anderson prescription reproduces the BdG results very well.}
\end{figure}

The accuracy of the Anderson results can be seen more closely 
in Fig.~2, where the cross sections of $\Delta_i$ versus $i$
in Fig.~1 for (a) $n = 0.2$ and (b) $n = 1.0$ are shown.
It is clear that the essential features of the
BdG results are reproduced in the Anderson approach.

To examine the effect of impurities, we show in Fig. 3 
$\Delta_i$ as a function of $i$, for $N=32$ 
with an impurity at the central site
with varying energy (both negative and positive). 
We have used PBC and 
an intermediate coupling strength, $|U| = 2\,t$,
at electron density $n = 0.9$.
We have intentionally stayed away from half-filling, 
at which
the ground state is not a superconducting state, but a charge density wave.
One may recall that with periodic boundary conditions and with no
impurities,
the ground state for the attractive Hubbard model at half-filling
is doubly degenerate: both superconducting and charge density wave
solutions coexist at this point. However, the presence of an impurity tilts 
the balance in favour of the charge density wave, and the BdG equations
converge to a solution in which the pair potential, $\Delta_i$, is
identically zero at all sites. The Hartree potential, $V_i$, 
on the other hand, oscillates as a function of site position.
The Anderson prescription
is unable to reproduce this (correct) feature at half filling,
and gives a superconducting solution with nonzero gap parameters.
Also 
if the self-consistency of the Hartree potential is neglected in
the BdG equations, this physics is missed, and the ensuing BdG result is
similar to the Anderson solution. 

\begin{figure}
\psss{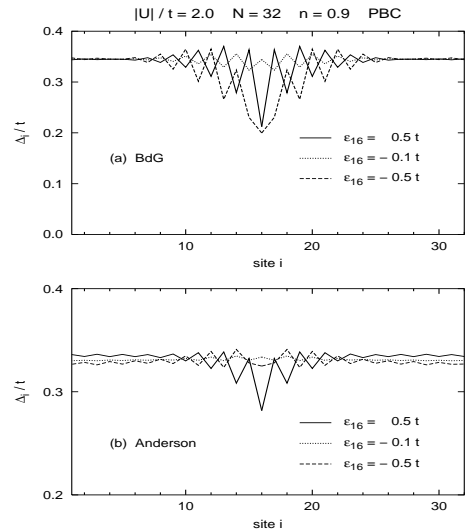}
\caption{Gap parameter $\Delta_i$ as a function of site number $i$, for
a chain of $N=32$ with PBC and an impurity at site 16 with varying potential
energy, from the (a) BdG and (b) Anderson equations.
Note that in (b) the scale of the gap is twice the scale in (b).}
\end{figure}

Returning to Fig.~3 we have plotted the BdG results in Fig.~3(a), while 
in Fig.~3(b) we show the results from the Anderson equations. 
The gap is suppressed at the site with a positive impurity potential
(as is the site density $n_i$)
and exhibits `Friedel-like' oscillations around it.
The Anderson prescription captures this behaviour qualitatively,
while it tends to underestimate the amplitudes of the oscillations 
(compare the solid curves in Fig.~3(a) and (b) for $\epsilon_{16}=0.5\,t$,
and note the magnified scale in the latter graph).
The Anderson results become better
for stronger impurity potentials and for electron density $n$ 
further away from half filling.
For a negative impurity potential, when the potential strength is very weak,
$\Delta_i$ (and $n_i$) has a peak at the impurity site, as can be seen
in Fig.~3(a) for $\epsilon_{16}=-\,0.1\,t$ (the dotted curve).
Though smaller in scale, the Anderson result in Fig.~3(b)
has similar behaviour.
As the potential strength increases, however, an attractive impurity
tends to break the pairing, and 
suppresses the gap not only at the impurity site
but also at surrounding sites
(see the dashed curve for $\epsilon_{16}=-\,0.5\,t$ in Fig.~3(a)).
In such cases, the Anderson method overestimates the gap parameter 
around the impurity site.
This can be seen in Fig.~3(b), where the gap has
the correct oscillating pattern, but 
with much smaller amplitudes.

\begin{figure}
\psss{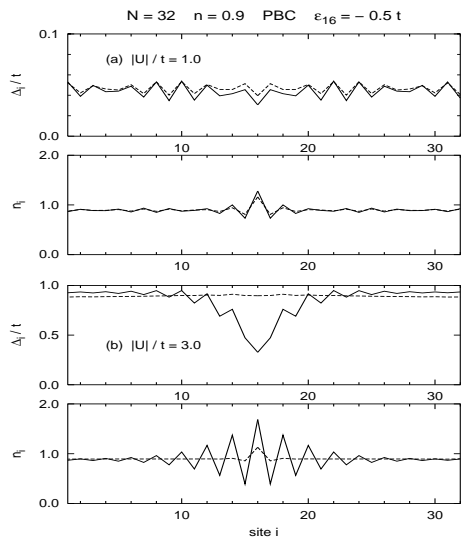}
\caption{The $\Delta_i$ as a function of $i$, for
a chain of $N=32$ with PBC and $\epsilon_{16}=-\,0.5\,t$,
for (a) $|U| = 1\,t$ and (a) $|U| = 3\,t$.
The BdG and Anderson results are plotted in solid and dashed curves,
respectively.}
\end{figure}

We study the attractive impurity case
further in Fig.~4, where 
we show $\Delta_i$ and $n_i$ versus $i$ for $N=32$ with PBC
and $\epsilon_{16}=-\,0.5\,t$, for (a) $|U| = 1\,t$ and (b) $|U| = 3\,t$.
When the coupling is
weaker than or comparable to the impurity potential, the Anderson approach
captures the main features of the gap parameter around the impurity.
This is the case for $|U| = 1\,t$ in Fig.~4(a), and indeed the Anderson
results show excellent agreement with the BdG results.
As the coupling becomes stronger compared to the impurity potential,
the impact of the impurity becomes more drastic,
even with relatively weak strength.
The gap is more suppressed around the impurity, and
the density distribution exhibits `Friedel-like' oscillations more enhanced.
This can be seen for $|U| = 3\,t$ in Fig.~4(b) (solid curves).
On the other hand, the Anderson method yields a gap parameter that is more 
uniform as a function of site position, and does not reproduce the correct
oscillations in the site densities.

In conclusion, we have formulated the BdG equations for a tight-binding
model with an on-site attractive interaction. We have retained the
self-consistent Hartree potential in the BdG equations, and found that
a single impurity breaks the superconducting/charge density wave degeneracy
which would otherwise exist at half-filling in this model. We have also
formulated the prescription set out by Anderson, without 
impurity-averaging, and found good qualitative agreement with
the BdG results. 

To our knowledge, the spatial dependence of the order parameter and
the electron density has not been previously explored in detail within the
Anderson prescription. We have found, somewhat to our surprise, good
agreement with results from the BdG formalism, for surfaces and single
impurities (aside from the weak scattering limit). Throughout this study,
it is important to note that in the vicinity of surfaces or impurities,
``charge ordered'' states and superconductivity in general coexist (at
the mean field level).
In future work we will examine various correlation functions
and the local density of states, the latter of which has been \cite{hudson99}
and will continue to be measured using scanning tunneling microscopy. 
 
{\it Acknowledgements} We thank Jorge Hirsch for suggesting the weak
impurity potential regime as one in which the Anderson prescription
should break down.  We also thank Kamran Kaveh for stimulating discussions.
Calculations were performed on the 64-node SGI
parallel processor at the University of Alberta.
This research was supported by the Avadh Bhatia Fellowship and by the
Natural Sciences and Engineering Research Council of Canada and
the Canadian Institute for Advanced Research.
One of us (F.M.) acknowledges the hospitality of the Aspen Center for 
Physics, where some of this work was performed.


\vskip-0.4cm

\begin{thebibliography} {999}

\bibitem{anderson59}
P. W. Anderson, J. Phys. Chem. Solids {\bf 11}, 26 (1959).

\bibitem{abrikosov59}
A.A. Abrikosov and L.P. Gor'kov, Sov. Phys. JETP {\bf 8}, 1090 (1959). 

\bibitem{physics_today}
See G.P. Collins, {\it in Physics Today, Search and Discovery}, p.17, 1993,
and M.F. Crommie, C.P. Lutz, and D.M. Eigler, Science {\bf 262}, 218 (1993).

\bibitem{rainerbook}
{\it Quasiclassical Methods in Superconductivity and Superfluidity},
edited by D. Rainer and J.A. Sauls, 1996.

\bibitem{degennes66}
P. G. de Gennes, {\it Superconductivity of Metals and Alloys} 
(W.A. Benjamin, Inc. New York, 1966).

\bibitem{randeria98}
A. Ghosal, M. Randeria, and N. Trivedi, Phys. Rev. Lett. {\bf 81}, 3940 (1998).

\bibitem{hudson99}
E.W. Hudson, S.H. Pan, A.K. Gupta, K.-W. Ng, and J.C. Davis, 
Science {\bf 285}, 88 (1999).

\bibitem{hirsch92} For a tight-binding formulation of the BdG equations,
see J.E. Hirsch, Physica C {\bf 194} 119 (1992). Note that he excluded
the Hartree-Fock potential.

\bibitem{tanaka00}
K. Tanaka and F. Marsiglio, Phys. Lett. A {\bf 265}, 133 (2000).

\bibitem{trugman} We thank Stuart Trugman for alerting us to this analytical
solution.

\bibitem{chain} In this work we use one dimensional chains, partly because
real space results are more easily illustrated, and partly because for the
attractive Hubbard model, dimensionality should not play a very important
role (at the mean field level), so that results presented here are
generally applicable to two and three dimensions.

\end{thebibliography}
\end{document}